\documentclass[aps,pra,notitlepage,twocolumn,superscriptaddress]{revtex4-1}
\usepackage[T1]{fontenc}
\usepackage{cancel}
\usepackage{amsmath,amssymb,amsfonts,amsthm,multirow,xcolor,graphicx,float,siunitx,dsfont,array}
\usepackage[colorlinks,breaklinks]{hyperref}
\usepackage[caption=false]{subfig}
\definecolor{darkred}{rgb}{0.5,0,0}
\definecolor{darkgreen}{rgb}{0,0.5,0}
\definecolor{darkblue}{rgb}{0,0,0.5}
\hypersetup{linkcolor=darkred,citecolor=darkgreen,urlcolor=darkblue}
\def\ket#1{|#1\rangle}

\def\ketbra#1{|#1\rangle\langle#1|}
\def\av#1{\langle#1\rangle}

\def\mub#1#2{\ket{\varphi_{#2}^{#1}}}

\renewcommand{\leq}{\leqslant}
\renewcommand{\geq}{\geqslant}

\newcommand{\rhoAB}{\ensuremath{\rho^{\text{AB}}}}
\newcommand{\betaLHS}{\ensuremath{\beta^{\text{LHS}}}}
\newcommand{\tbetaLHS}{\ensuremath{\tbeta^{\text{LHS}}}}
\newcommand{\ta}{\ensuremath{\tilde a}}
\newcommand{\tA}{\ensuremath{\tilde A}}
\newcommand{\tbeta}{\ensuremath{\tilde\beta}}
\newcommand{\tF}{\ensuremath{\tilde F}}
\newcommand{\tmu}{\ensuremath{\tilde\mu}}
\newcommand{\tx}{\ensuremath{\tilde x}}

\DeclareMathOperator{\tr}{tr}
\usepackage{tabularx}
\usepackage[framemethod=tikz]{mdframed}
\definecolor{mycolor}{rgb}{0.122, 0.435, 0.698}
\definecolor{mycolor2}{RGB}{166,83,154}
\newmdenv[innerlinewidth=0.5pt,roundcorner=4pt,linecolor=mycolor,innerleftmargin=6pt, innerrightmargin=6pt,innertopmargin=6pt,innerbottommargin=6pt]{mybox}
\usepackage{paracol}
\usepackage{tcolorbox}
\tcbset{colback=mycolor2!5,colframe=mycolor2,fonttitle=\bfseries, float=htb}
\newtcolorbox[blend into=figures]{boxfigure}[3][]{float*=ht,width=\textwidth,lower separated=false,center upper,title={#2},label=fig:#3,#1}
\newtcolorbox[blend into=figures]{smallboxfigure}[3][]{float=ht,lower separated=false, blend before title=colon hang,title={#2},label=fig:#3,#1}
\newtcolorbox{smallbox}[3][]{float=ht,lower separated=false,blend before title=colon hang,title={#2},label=fig:#3,#1}
\newtcolorbox[blend into=tables]{smallboxtable}[3][]{float=tb,lower separated=false, blend before title=colon hang,title={#2},subtitle style={boxrule=0.4pt,
colback=mycolor2!50!red!25!mycolor2},label=table:#3,#1}
\begin{document}
\title{Noise-Robust and Loss-Tolerant Quantum Steering with Qudits }
\author{Vatshal Srivastav}
\email[Email address: ]{vs54@hw.ac.uk}
\affiliation{Institute of Photonics and Quantum Sciences, Heriot-Watt University, Edinburgh, UK}
\author{Natalia Herrera Valencia}
\affiliation{Institute of Photonics and Quantum Sciences, Heriot-Watt University, Edinburgh, UK}
\author{Will McCutcheon}
\affiliation{Institute of Photonics and Quantum Sciences, Heriot-Watt University, Edinburgh, UK}
\author{Saroch Leedumrongwatthanakun}
\affiliation{Institute of Photonics and Quantum Sciences, Heriot-Watt University, Edinburgh, UK}
\author{S{\'e}bastien Designolle}
\affiliation{Department of Applied Physics, University of Geneva, 1211 Geneva, Switzerland}
\author{Roope Uola}
\affiliation{Department of Applied Physics, University of Geneva, 1211 Geneva, Switzerland}
\author{Nicolas Brunner}
\affiliation{Department of Applied Physics, University of Geneva, 1211 Geneva, Switzerland}
\author{Mehul Malik}
\email[Email address: ]{m.malik@hw.ac.uk}
\affiliation{Institute of Photonics and Quantum Sciences, Heriot-Watt University, Edinburgh, UK}

\date{\today}
\begin{abstract}
A primary requirement for a robust and unconditionally secure quantum network is the establishment of quantum nonlocal correlations over a realistic channel. 
While loophole-free tests of Bell nonlocality allow for entanglement certification in such a device-independent setting, they are extremely sensitive to loss and noise, which naturally arise in any practical communication scenario. 
Quantum steering relaxes the strict technological constraints of Bell nonlocality by re-framing it in an asymmetric manner, thus providing the basis for one-sided device-independent quantum networks that can operate under realistic conditions.
Here we introduce a noise-robust and loss-tolerant test of quantum steering designed for single detector measurements that harnesses the advantages of high-dimensional entanglement.
We showcase the improvements over qubit-based systems by experimentally demonstrating detection loophole-free quantum steering in 53 dimensions through simultaneous loss and noise conditions corresponding to 14.2~dB loss equivalent to 79~km of telecommunication fibre, and $36\%$ of white noise.
We go on to show how the use of high dimensions counter-intuitively leads to a dramatic reduction in total measurement time, enabling a quantum steering violation almost two orders of magnitude faster obtained by simply doubling the Hilbert space dimension. 
By surpassing the constraints imposed upon the device-independent distribution of entanglement, our loss-tolerant, noise-robust, and resource-efficient demonstration of quantum steering proves itself a critical ingredient for making device-independent quantum communication over long distances a reality.
\end{abstract}

\maketitle
\section{Introduction}
\begin{figure}[ht!]
  \centering
  \includegraphics[width=0.97\columnwidth]{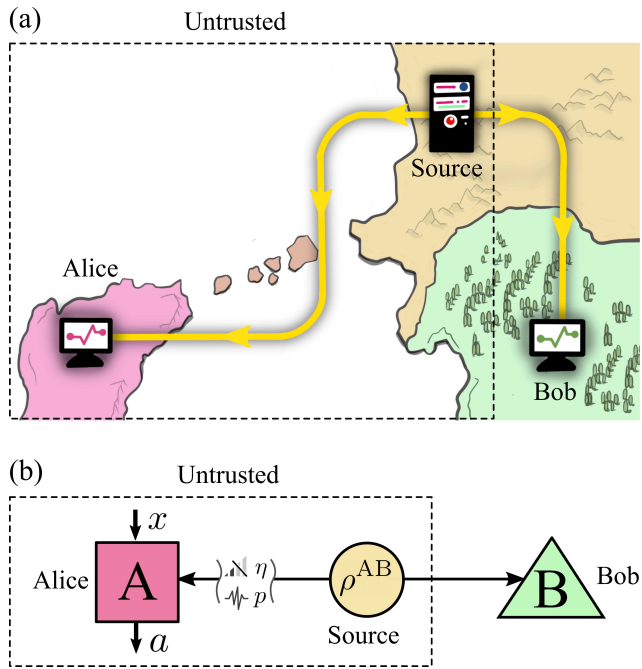}
  \caption{\textbf{One-sided device-independent (1SDI) quantum communication scenario.} The illustration in (a) depicts an example of a fibre-based 1SDI quantum network between two distant stations, Alice and Bob. A source distributes entangled photon pairs to Alice and Bob via optical fibre links. In the SDI scenario, the source, Alice's base station, and the fibre link between them are untrusted, for example due to their geographic location or compromised devices.
  On the other hand, Bob's station is trusted and hence imperfections on his side are ignored. 
  A suitable strategy to certify entanglement in this scenario is given by quantum steering (b), where a bipartite state $\rhoAB$ is shared between Alice and Bob. 
  Alice's measurements and the channel between her and the source are affected by loss ($\eta$) and noise ($p$) and considered to be untrusted. By performing local measurements $x$ with outcomes $a$, Alice conditions the shared state $\rhoAB$ to an assemblage $\sigma_{a|x}$ at Bob. Bob can perform suitable measurements on his shared state and check if the state is steerable (thus entangled) by violating a steering inequality.}
  \label{fig:basic_principle}
\end{figure}
In today's digital landscape riddled with threats such as cyber-attacks and information leaks, the advent of secure quantum communication has strongly impacted modern technological progress.
Intense research advances have been carried out in the past two decades to achieve a secure and robust implementation of quantum communication between two distant parties \cite{Gisin2007,Diamanti2016,Krenn2016,Feihu20,Scarani09}.
The ultimate form of security is provided in the scenario when the two parties, Alice and Bob, can verify entanglement between them in a device-independent (DI) manner~\cite{Acin:2007db}, i.e., without requiring any trust in their devices or the channels themselves.
A requirement for this form of entanglement distribution is a test of quantum nonlocality, such as the loophole-free violation of a Bell inequality, which has been demonstrated over short distance scales of up to a kilometre \cite{Giustina2015,Shalm2015,Hensen2015}.
However, inevitable loss due to propagation and environmental noise restrict the maximum distance over which entanglement can be certified in a DI manner, making such protocols vulnerable to attacks associated with the detection loophole (where an eavesdropper could exploit an unjustified fair-sampling assumption)~\cite{Pearle70,Gisin1999}. 
In other words, we must associate our inability to measure every photon that was created with the actions of a malicious eavesdropper ``listening in'' on the quantum conversation.
Closing the detection loophole is technologically demanding, normally requiring extremely high overall system detection efficiencies,  which naturally imposes practical limitations over realistic long-distance channels.

Quantum steering is an alternative scenario that relaxes the rigid technical requirements of device-independent entanglement certification.
Here, one can assume that a trusted measurement device exists only on one side~\cite{WJD07,CS16b,UCN+20}.
In this asymmetric one-sided device-independent (1SDI) setting, entanglement is certified when the untrusted party (say, Alice) is able to condition or ``steer'' the state of the trusted party (Bob) through her local measurements.
Note that the detection loophole is a threat only on Alice's side due to her untrusted measurement apparatus, as well as any loss or noise introduced in the untrusted channel. 
Since Bob's measurement devices are trusted, he is exempted from the loophole. Fig.~\ref{fig:basic_principle}(a) illustrates an example of such a scenario, where Alice and the entanglement source are located in different (untrusted) geographic locations, and are connected by an inaccessible (untrusted) undersea fibre-optic channel.

The experimental detection of steering is conveniently achieved via the violation of so-called steering inequalities \cite{Cavalcanti2009,Saunders2010}. Recently, such violations were reported experimentally, with the detection loophole closed, using qubit entanglement \cite{Wittmann12,BES+12,Smith2012}. By increasing the number of measurement settings used by both parties, the threshold heralding efficiency required to demonstrate EPR steering with qubit entanglement can be lowered arbitrarily, in the limit of infinitely many measurement settings.
However, this is only possible if Alice and Bob share a high-quality qubit entangled state, with higher loss requiring increasingly higher state qualities~\cite{BES+12}.
In addition to loss, any realistic quantum communication system will be prone to sources of noise such as dark and background counts, co-propagating classical signals, and imperfect measurement devices. 
Furthermore, performing a large number of measurements can be impractical, leading to extremely long measurement times in an already lossy scenario~\cite{BES+12, WSC+18}.

It has been established that high-dimensional entanglement can overcome several limitations of qubit-entangled systems \cite{Cozzolino:2019ct,Erhard:2019ux}. Such entangled ``qudits'' can exhibit stronger correlations than qubit entanglement \cite{Brunner2008,Des21}, and can tolerate lower heralding efficiency thresholds for tests of nonlocality \cite{Vertesi:2010bq}. Qudits also offer advantages in quantum key distribution \cite{Tittel2000,Cerf02,Malik:2014ht}, leading to higher key rates and robustness to noise \cite{Sheridan2010, Nunn:13,Zhang14,Mirhosseini:2015fy}.
Notably, the large dimensionality of photonic platforms has enabled entanglement distribution with greater noise resistance~\cite{EckerHuber2019,Zhu:2019tb, DSU+21} and higher information encoding capacities~\cite{Valencia:2020gx,Cao:2020,Erhard:2017gl,Steinlechner:2017bw,Mafu:2013jk} than qubits.
These merits of high-dimensional entanglement make it a strong contender for the realisation of device-independent quantum communication protocols~\cite{Acin:2007db}; see also \cite{HuberPaw2013}. 
However, many practical considerations have hindered their adoption---general multi-outcome measurements of high-dimensional quantum states of light are notoriously difficult to realise, for, e.g., requiring arrays of cascaded, unbalanced interferometers \cite{Takesue:2017} or complex spatial mode transformation devices \cite{Fontaine:2019ja,SG2022}. 
In addition, they are prone to impractically long measurement times and suffer from loss and noise due to the lack of ideal multi-outcome detectors.

Here, we overcome many of the challenges associated with high-dimensional photonic systems through simultaneous advances in theory and experiment, allowing us to demonstrate quantum steering with the detection loophole closed under extreme conditions of loss and noise.
First, we formalise a set of linear steering inequalities requiring only a single detector at each party, unlike standard multi-outcome measurements that require $d$ detectors for measuring qudits.
These inequalities are especially formalised for high-dimensions by \textit{binarising} projective measurements onto mutually unbiased bases (MUBs)~\cite{WF89}, and exhibit the same loss-tolerance and noise-robustness as their counterparts designed for multi-outcome detectors~\cite{SGMS16}. 
In contrast however, these provide the strong advantages of needing significantly fewer technological resources and being free from strong assumptions on detector noise, such as background or accidental count subtraction.
We violate these steering inequalities experimentally with photon pairs entangled in their discrete transverse position-momentum, in the presence of noise and at a record-low heralding efficiency of $0.038\pm 0.001 $ ($\sim 14.2$~dB), which is equivalent to the optical loss of a 79~km-long telecommunications fibre \cite{fibreloss}.

Despite the number of single-outcome measurements scaling with dimension $O(d^2)$, we show that the total measurement time in large $d$ can be reduced considerably thanks to the high statistical significance of a violation in high dimensions for a fixed number of counts. 
We experimentally verify this by comparing the violations at two different dimensions ($d = 23, 41$) taken at two different acquisition times under a fixed channel loss ($\sim 12.1$~dB). 
As a result, we are able to drastically reduce the total measurement time from $2.53$~hours for $d =23$, to $2.5$~mins for $d = 41$. In both dimensions, we violate the steering inequality by ten standard deviations. Below, we elaborate on the theoretical formulation of our binarised linear steering inequality, followed by a detailed discussion of the experimental implementation and results.

\section{Theory}
\begin{figure}[t!]
  \centering
  \includegraphics[width=\columnwidth]{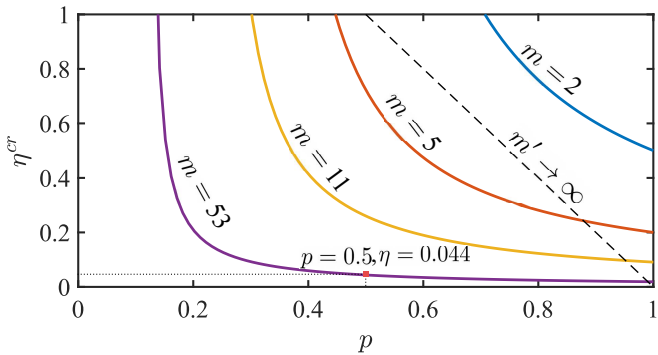}
  \caption{\textbf{Noise robustness and loss tolerance.} Dependence of the critical heralding efficiency $\eta^{cr}$ for Alice with respect to the noise parameter $p$ in the bipartite shared state $\rhoAB(p)$ when using $m = d$ MUB measurement settings to demonstrate EPR steering in dimension $d$.
  By increasing the number of measurement settings $m$ (or increasing the dimensions $d$), the critical efficiency $\eta^{cr}$ decreases substantially, even at significant noise levels ($p<1$). For example, with $m=53$ MUB settings, one can tolerate a heralding efficiency as low as $\eta=0.044$ and a noise parameter $p=0.5$ (equal mixture of maximally entangled state and white noise).
    In contrast, demonstrating steering with qubit entanglement ($d=2$) requires significantly higher heralding efficiencies $\eta_{cr}$ and a noise parameter $p>0.5$ in the impractical limit of infinite measurement settings ($m'\rightarrow\infty$, dashed line)~\cite{BES+12}. Thus, high-dimensional entanglement enables us to access regimes of noise and loss that are inaccessible by qubit entanglement, even in the best-case scenario.}
  \label{fig:pvsetamod}
\end{figure}
%%%%%%%%%%%%%%%%%%%%%%%%%%%%%%%%%%%%%%%%%%%%%%%%%
%%%%%%%%% Setup Figure %%%%%%%%%%%%%
\begin{figure*}[ht!]
  \centering
  \includegraphics[width=1\textwidth]{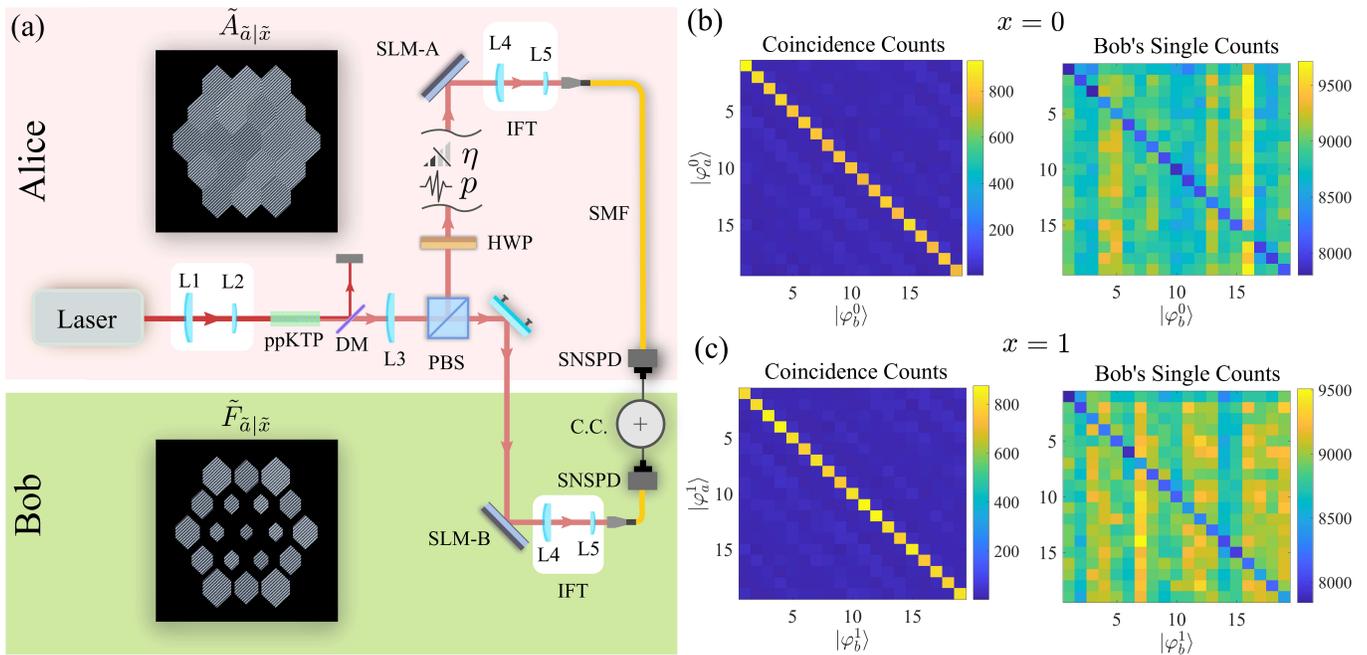}
  \caption{\textbf{Experimental setup.} (a) A Ti:Sapphire pulsed laser at 775~nm is used to pump a nonlinear ppKTP crystal to generate a pair of photons entangled in their transverse position-momentum via Type-II spontaneous-parametric-down-conversion (SPDC).
  The pump photons are filtered by a dichroic mirror (DM) and the downconverted photons are separated with a polarising-beam-splitter (PBS). 
  One photon from the entangled photon pair is sent to the untrusted party, Alice, who generates holograms on a spatial light modulator (SLM-A) to perform the projective measurements $\tA_{\ta|\tx}$, see Eq.~\eqref{eqn:measall}. 
  Bob receives the other photon along with the outcome $\ta$ from Alice, forming the assemblage $\sigma_{\ta|\tx}$. 
  He then performs the projective measurements $\tF_{\ta|\tx}$ (Eq.~\eqref{eqn:bobbin}) according to the steering inequality defined in Eq.~\eqref{eqn:generalineq}.
  Each of the photons in the selected mode are collected by a combination of telescope lenses (L4 and L5) into single-mode fibres (SMF) and are detected by superconducting nanowire detectors (SNSPDs). 
  Coincident detection events corresponding to joint two-photon measurements are recorded by a coincidence counting logic (CC) with a coincidence window of 0.2~ns. 
  (b) and (c) Experimental data showing coincidence counts between Alice and Bob and exclusive single counts measured on Bob's side in dimension $d=19$ using MUB measurement settings $x = 0,1$. 
  }
  \label{fig:expsetup}
\end{figure*}
%%%%%%%%%%%%%%%%%%%%%%%%%%%%%%%%%%%
In quantum steering, the untrusted party Alice conditions the shared bipartite state $\rhoAB$ through her measurement operators $\{A_{a|x}\}$, where $x$ denotes her choice of measurement and $a$ her outcome. 
By doing so, she creates the conditional (non-normalised) states ${\sigma_{a|x}=  \tr_A\left[(A_{a|x}\otimes\openone_B)\rhoAB\right]}$, also known as an assemblage.
Since the trusted party Bob has access to the assemblage, he can check if $\sigma_{a|x}$ can be produced without the use of entanglement via a so-called local hidden state (LHS) model~\cite{WJD07}. 
A steering inequality \cite{Cavalcanti2009,Saunders2010} allows Bob to detect assemblages that do not follow any LHS model and thus demonstrate steering. 
Formally, it consists of a set of (unnormalised) measurements $\{F_{a|x}\}$ on Bob's side and a value $\betaLHS$ such that, for all unsteerable assemblages we have that

\begin{equation}
  \beta\equiv\sum_{a,x}\tr(F_{a|x}\sigma_{a|x}) \leq \betaLHS,\label{eqn:generalineq}
\end{equation} 
where $\betaLHS$ is the maximum value of the functional $\beta$ for any unsteerable, i.e., LHS assemblage. When using entanglement and appropriate measurements, higher functional values, $\beta_Q$ can be obtained. Hence steering is demonstrated whenever the above inequality in Eq.~\eqref{eqn:generalineq} is violated, that is, when $\beta_Q>\betaLHS$.

Here, we formalise a set of linear steering inequalities designed especially for single-outcome projective measurements, which are particularly suitable for single-photon detection systems that are widely implemented on photonic platforms.
First, we begin with a steering inequality in which Alice and Bob measure $d$-outcome projective measurements in a given $d$-dimensional mutually unbiased basis (MUB). Alice (Bob) measures $A_{a|x}$ ($A_{a|x}^T$) with an outcome $a=0,..,d-1$ in a basis given by $x = 0,\dots,m$ with $m\leq d$, see Eq.~\eqref{eqn:MUBformula}. We then binarise these measurements such that Alice's measurements become

\begin{equation}
  \tA_{\ta|\tx}(\eta)=
  \begin{cases}
    \eta A_{a|x}&\text{if }\ta=1\\
    \openone-\eta A_{a|x}&\text{if }\ta=0,
  \end{cases}
  \label{eqn:measall}
\end{equation}
where $\eta$ is the one-sided heralding efficiency at Alice.
The one-sided heralding efficiency is defined as the probability that detecting a photon at the trusted party (Bob) heralds the presence of a photon at the untrusted party (Alice). %define heralding efficiency
Note that the binarised measurements $\tA_{\ta|\tx}$ are labelled by new indices $\ta=0,1$, corresponding respectively to a no-click/click event at the detector and $\tx = (a,x)$, reflecting the new set of measurement settings that includes every projector from each basis.
Bob's measurements, which are used to evaluate the steering inequality, are defined in a similar fashion 

\begin{align}\label{eqn:bobbin}
  \tF_{\ta|\tx}=
  \begin{cases}
    A_{a|x}^T&\text{if }\ta=1\\
    c\,(\openone-A_{a|x}^T)&\text{if }\ta=0,
  \end{cases}
\end{align}
where $c = 1/(\sqrt{d}-1)$ is a constant chosen to facilitate the derivation of a closed form expression for an upper bound, $\tbeta$, on the corresponding LHS bound, given by $ \tbetaLHS\leq1+m(\sqrt{d}+1)\equiv \tbeta$, see Eq.~\eqref{eqn:desiredbLHS} in Appendix~\ref{methods:betaLHSbin}.
Therefore, obtaining a value of the steering inequality larger than $\tbeta$ demonstrates steering.

Second, in order to study the effects of noise suffered by the shared state $\rhoAB$, one can consider the isotropic state that includes a fraction of added white noise, specifically, $\rhoAB(p)\equiv p\ketbra{\phi_d} + (1-p)\openone_{d^2}/d^2$, where ${\ket{\phi_d}\equiv\sum_{i=0}^{d-1}\ket{ii}/\sqrt{d}}$ is the maximally entangled state in dimension $d$ and $p\in [0,1]$ is the noise mixing parameter. Finally, we use Eq.~\eqref{eqn:generalineq} to evaluate the functional $\beta^Q$ in terms of one-sided heralding efficiency $\eta$ and mixing parameter $p$, see Eq.~\eqref{eqn:betaall}.
To check the violation of the steering inequality, we calculate the difference between $\beta^Q(\eta, p)$ and $\tbeta$ (upper bound on $\tbetaLHS$), which is given as,

\begin{align}\label{eqn:deltabeta}
   \Delta \beta  = \eta m \left(p-\frac{1-p}{\sqrt{d}}\right)-1. 
\end{align}

The condition $\Delta \beta  > 0$ must be satisfied in order to demonstrate steering, which also leads to a critical value $\eta^{cr}$ for the one-sided heralding efficiency $\eta$.
The critical heralding efficiency $\eta^{cr}$ generally depends on the noise parameter $p$, the number of MUB measurement settings $m$, and the dimension of the Hilbert space $d$ (see Eq.~\eqref{eqn:generaletacr}).
In the noise-free case ($p = 1$), the critical one-sided heralding efficiency can be reduced to $\eta^{cr}  = 1/m$, the lowest possible value \cite{Pironio2003}, which can also be obtained with qubits~\cite{BES+12}.
However, any experiment features noise, i.e., with a mixing parameter $ p<1$, originating from various technical imperfections ranging from detector dark counts or multi-pair emission to misalignments in the system \cite{Zhu:2019tb}. 
As the amount of noise increases, it can be shown that the value of $\eta^{cr}$ required to demonstrate steering increases, see Eq.~\eqref{eqn:generaletacr}. 
For qubits ($d = 2$), the required critical efficiency $\eta^{cr}$ to demonstrate steering reaches unity when the noise parameter $p \sim 0.71$ for $m=2$ measurements, while steering is impossible for $p\leq 1/2$ \cite{WJD07,Werner1989}.

In contrast, Fig.~\ref{fig:pvsetamod} shows that by increasing the number of MUB measurement settings $m$, which is only possible in the qudit regime ($d>2$), one can still demonstrate steering in the presence of substantial loss and noise in the channel as compared to qubit-based systems. For example, using 53 MUB settings, one can tolerate a heralding efficiency as low as $\eta=0.044$ and a noise parameter of $p=0.5$.
This enables one to find the perfect trade-off between loss in the channel and noise in the system.
Additionally, with our binarised steering inequalities, projective measurements with only two outcomes (photon detected or not detected) not only show the same loss-tolerance and noise-robustness as their multi-outcome counterparts for steering~\cite{SGMS16}, but are also more feasible to implement experimentally with two single-click photon detectors that are typically available in every photonics laboratory.

%%%%%%%%%%%%%%%%% Results%%%%%%%%%%%%%%%%%%%%%%%
\begin{figure}[t!]
  \centering
  \includegraphics[width=1\columnwidth]{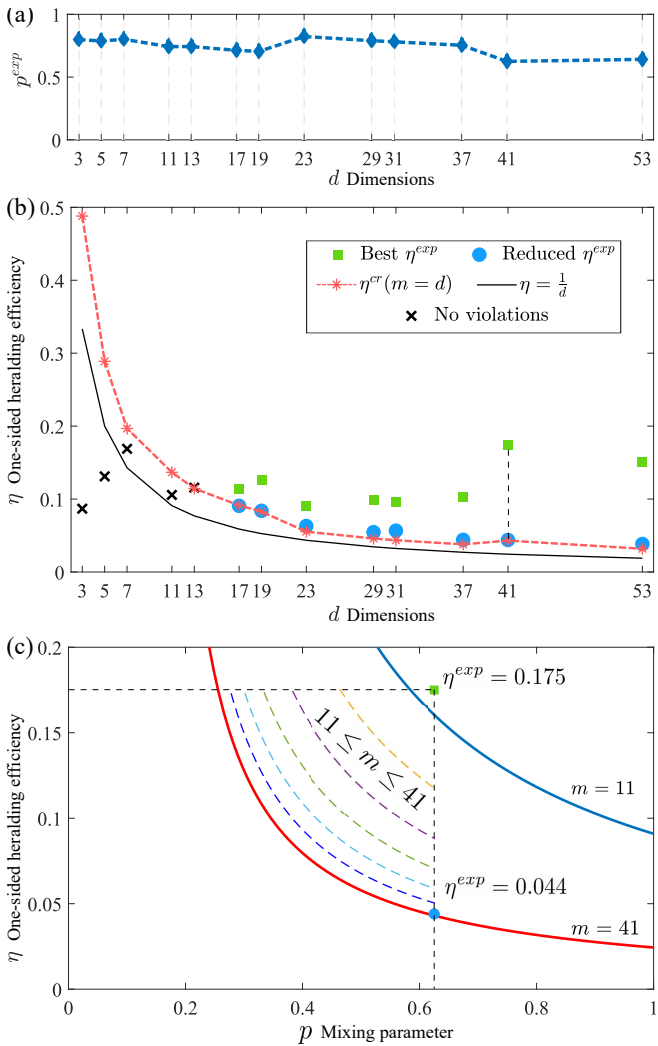}
  \caption{\textbf{Experimental results.} (a) The experimental noise mixing parameter $p^{exp}$ as a function of state dimension $d$, corresponding to a $(1-p)$ fraction of white noise present in the state. (b) The critical one-sided heralding efficiency required to demonstrate steering decreases as a function of $d$, as shown for the noise-free case ($\eta$ with $p=1$, solid black curve) and non-ideal state quality ($\eta^{cr}$ with $p^{exp}<1$, dashed red curve). In dimensions $3\leq d<17$ (black crosses), our experimental one-sided heralding efficiency is under the critical value ($\eta^{exp} <  \eta^{cr}(m=d)$), thus the state does not demonstrate steering.
  In dimensions $d\geq 17$, we are able to demonstrate steering with similarly low heralding efficiencies $\eta^{exp}$ (green squares) by using $m = d$ MUB settings. The experimental heralding efficiency can be lowered further (blue dots) by adding loss in Alice's channel, allowing us to demonstrate EPR steering in $d=53$ with a record-low $\eta^{exp}=0.038$ and a noise parameter $p^{exp}=0.641$.
  (c) To showcase the loss-tolerance of our binarised steering inequality, we focus on the case at $d=41$. At $\eta^{exp}= 0.175$, one can demonstrate steering with fewer measurement settings ($m=11$).
  However, when Alice's channel efficiency is reduced by a factor of four to $\eta^{exp}= 0.044$, EPR steering can still be demonstrated by using more measurement settings ($m=41$).}
  \label{fig:results}
\end{figure}
%%%%%%%%%%%%%%%%%%%%%%%%%%%%%%%%%%%%%%%%

\section{Experiment}
To experimentally demonstrate noise-robust EPR steering with the detection loophole closed, we use pairs of photons entangled in their discrete transverse position-momentum degree-of-freedom, also known as pixel entanglement \cite{Srivastav2021,HerreraValencia2020highdimensional}.
As shown in Fig.~\ref{fig:expsetup}(a), two spatially entangled photons at 1550~nm are generated in a 5~mm nonlinear ppKTP crystal through the process of Type-II spontaneous parametric down conversion (SPDC) using a Ti:Sapphire femtosecond pulsed laser with 500~mW average power.
After being separated by a polarising beam-splitter (PBS), each photon from the entangled pair is directed to the two parties Alice and Bob, who can perform local projective measurements by using a spatial light modulator (SLM) to display tailored phase-only holograms. These holograms allow the measurement of a general superposition of spatial modes, i.e., a state from any MUB. 
Only photons carrying the correct states are efficiently coupled into single-mode fibres (SMF), which lead to two superconducting nanowire detectors (SNSPDs) connected to a coincidence logic (C.C.). This measurement system allows Alice and Bob to perform the local measurements in Eqs.~\eqref{eqn:measall} and \eqref{eqn:bobbin} and evaluate the elements of the functional in Eq.~\eqref{eqn:generalineq} in terms of the normalised coincidence counts between them, and of the normalised exclusive single counts measured by Bob, see Eq.~\eqref{eqn:functinalCCexcS} in Appendix~\ref{methods:experiments}. 

To close the detection loophole on Alice's side, her measurement basis is designed using hexagonal pixels of equal size with zero spacing between them. This enables Alice to maximise the detection efficiency of her SLM and channel. 
Bob's measurement basis design is informed by prior knowledge of the two-photon joint-transverse-momentum-amplitude (JTMA)~\cite{Srivastav2021}, which allows him to tailor his pixel mask in order to optimise the resultant one-sided heralding efficiency in the experiment $\eta^{exp}$ (see Appendix~\ref{methods:experiments}).
The choice of the phase-only pixel basis gives the added advantage that projective measurements in mutually unbiased bases provide the highest possible SLM heralding efficiency, since they do not require any amplitude modulation in their holograms \cite{Arrizon:07}.
Furthermore, the two-photon state encoded in the pixel basis can be designed to be close to a maximally entangled state in dimension $d$ owing to the knowledge of the JTMA of the generated two-photon state, which maximises its entanglement-of-formation~\cite{Srivastav2021,HerreraValencia2020highdimensional}.
In our experiment, we utilise the crosstalk between individual pixels as a reliable measure of the system noise parameter ($p^{exp}$, see Eq.~\eqref{eqn:mixingparameterexp}). 
This measure is valid because the amount of crosstalk between discrete spatial modes does not change substantially across the MUBs and thus behaves isotropically (see Appendix~\ref{methods:experiments} for more details).
\section{Results}

\begin{figure}[t!]
  \centering
  \includegraphics[width=1\columnwidth]{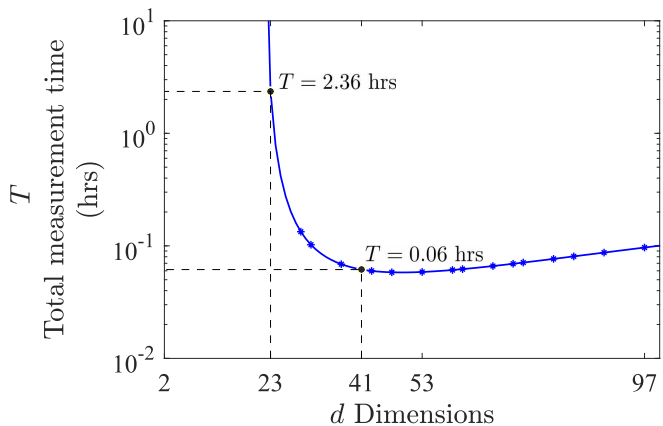}
  \caption{\textbf{Dependence of total measurement time ($T$) on dimension ($d$).} The total measurement time $T$ (on log scale) required to violate the steering inequality by 10 standard deviations can be calculated from Eq.~\eqref{eqn:timevsDim} as a function of prime dimension $d$ for $m = d$ measurement settings, for a fixed one-sided heralding efficiency $\eta = 0.062$ and noise parameter $p = 0.775$.
  The total time taken to obtain a steering violation using $m=23$ measurement settings in dimension $d=23$ is two orders of magnitude larger than that with $m = d = 41$.}
  \label{fig:timevsDim}
\end{figure}
%%%%%%%%%% Results%%%%%%%%%%%%%%%%%%%%%
In our experiment, we have demonstrated detection-loophole-free steering in up to dimension $d=53$ by performing binarised projective measurements in a number of MUBs ranging from $m_{\min}=12$ to ${m_{\max}=53}$. We exclude the computational (hex-pixel) basis in every dimension due to its much higher loss. The system noise parameter ($p^{exp}$) ranges from $p^{exp}=0.823$ to $0.625$ and is shown in Fig.~\ref{fig:results}(a) for every dimension.
In dimensions $d\geq17$ we introduce additional loss on Alice's channel by decreasing the diffraction efficiency of SLM-A and thus reducing the one-sided heralding efficiency $\eta^{exp}\rightarrow \eta^{cr}(m=d)$ until the state becomes just unsteerable (Fig.~\ref{fig:results}(b)).
For these reduced $\eta^{exp}$, we are able to demonstrate steering with $m =d$ MUB measurement settings.
Using $m=53$ MUB settings for example, we are able to tolerate a record-low one-sided heralding efficiency of $\eta^{exp}=0.038 \pm 0.001$, and violate our steering inequality by more than eight standard deviations. Note that this violation is obtained under noise conditions corresponding to $p^{exp}=0.641$. With lower noise, the one-sided heralding efficiency can be lowered further.

\begin{smallboxtable}[floatplacement=H]{Experimental measurement time $T$ required for a steering violation in two different dimensions $d$}{timeresults}
  \begin{center}
    \begin{tabularx}{\textwidth}{p{0.1\textwidth}p{0.35\textwidth}p{0.15\textwidth}p{0.2\textwidth}p{0.22\textwidth}}
      \hline
      $d$  & $\eta^{exp}$ & $t^{ac}$ & $T$ & $N$ \\
      \hline
      23  & $0.063 \pm 0.001$ & \SI{750}{\milli\second} & \SI{2.53}{\hour} & $\sim 10^4$ \\
      41  & $0.062\pm0.006$ & \SI{2.2}{\milli\second} & \SI{2.53}{\minute} & $\sim 70$
    \end{tabularx}
  \end{center}
  \tcblower
\footnotesize{The total measurement time required for a 10 standard deviation-violation of EPR steering in our experiment is 60 times lower in $d=41$ than in $d=23$. Here, $\eta^{exp}$ is the one-sided heralding efficiency, $t^{ac}$ is the acquisition time for a single measurement, and $N$ is the total number of singles counts measured per acquisition time window.
}
\end{smallboxtable}

We use our result in $d=41$ to highlight the tolerance to loss enabled by the use of high dimensions. Figure~\ref{fig:results}(c) shows the critical one-sided heralding efficiency $\eta^{cr}(m)$ required to demonstrate steering in $d=41$ using $m$ measurement settings, as a function of noise parameter $p$. As $m$ is increased, the critical efficiency at a given noise level can be significantly reduced. For example, at the fixed noise level $p^{exp} = 0.625$, we are able to demonstrate steering with a one-sided heralding efficiency of $\eta^{exp}=0.175$ using $m=11$ measurement settings. Note that this is possible as long as $\eta^{exp}$ satisfies $\eta^{exp} > \eta^{cr}(m)$.
As we increase the number of measurement settings up to $m = 41$, Alice's channel efficiency can be further reduced by a factor of four to $\eta^{exp}=0.044$ and still demonstrate quantum steering. This clearly demonstrates the increased robustness to loss enabled by high dimensions in detection-loophole-free steering violations. In general, the amount of loss and noise tolerated can be optimised by working along the critical one-sided heralding efficiency curves shown in Fig.~\ref{fig:results}(c) for a given number of measurement settings.
\begin{figure}[t!]
  \centering
  \includegraphics[width=1.0\columnwidth]{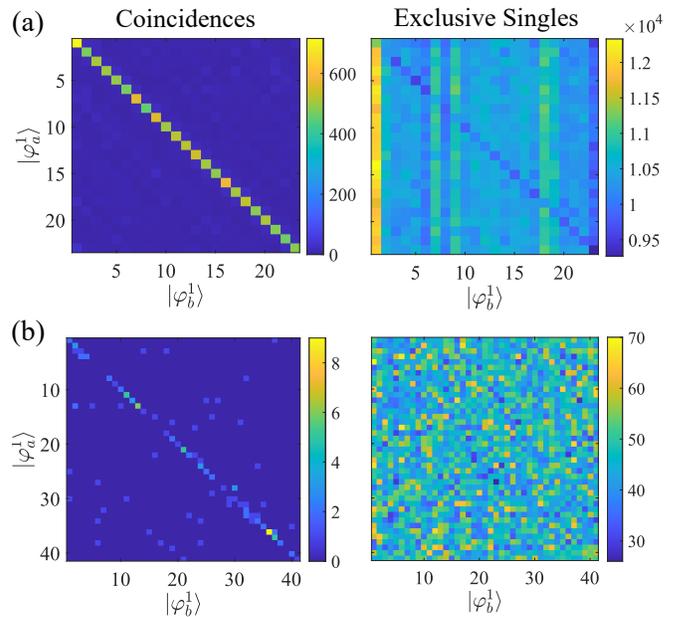}
  \caption{\textbf{Time-efficient steering in high dimensions.} The coincidence counts between Alice and Bob and exclusive single counts measured by Bob in measurement basis $x = 1$ in dimension (a) $d=23$ and (b) $d=41$ at a fixed one-sided heralding efficiency of $\eta^{exp}\sim0.062$, detected with two different acquisition times. Interestingly, even though the counts in $d=41$ are two orders of magnitude lower than in $d = 23$, we still violate the steering inequality with ten standard deviations in both cases.}
  \label{fig:timedata}
\end{figure}

A common problem encountered in experiments with qudits is that the total measurement time ($T$) increases drastically with the dimension of the Hilbert space, as the total number of single detector measurements scales as $md^2$. 
For our binarised steering inequality, we show that this is indeed not the case when one considers a steering violation within a fixed confidence interval.
Consider the example shown in Fig.~\ref{fig:timevsDim}, where for a given amount of noise and loss, a steering violation is only obtained in dimensions $d\geq 23$. As $d$ is increased, the total measurement time $T$ required to violate the steering inequality by ten standard deviations decreases dramatically till it attains a minimum, and then gradually increases again.
This allows us to minimise $T$ over larger dimensions at which the resilience against loss and noise is substantially high, and still demonstrate steering in a practical measurement time (see Appendix~\ref{methods:confidence} for a detailed derivation).
We verify this result in our experiment by comparing the experimental data taken at two different dimensions ($d_1=23, d_2=41$) using $m=d$ settings, at a fixed one-sided heralding efficiency $\eta^{exp}=0.062 \pm 0.006$ and noise parameter $p^{exp} = 0.775$, but with two very different acquisition times ($t_1^{ac}=$ \SI{750}{\milli\second}, $t^{ac}_2 =$ \SI{2.2}{\milli\second}) (see Table~\ref{table:timeresults}).
For $d_1 = 23$, the total measured singles counts per acquisition window are around $10^4$, while for $d_2 = 41$, counts are lowered over 100 times to about $70$. The coincidence counts in $d_2 = 41$ are just a few counts per second, on the order of the cross-talk (see Fig.~\ref{fig:timedata}).
While these two cases both demonstrate a violation of the steering inequality by ten standard deviations, $d_2=41$ achieves a substantial reduction in the total measurement time ($T_1 = \SI{2.53}{\hour}$ and $T_2 = \SI{2.53}{\minute}$).
Note that the response time of the spatial light modulators is excluded from the total measurement time $T$, as this is a technical limitation that can be addressed by using a fast phase modulation device such as a digital mirror device~\cite{Mirhosseini:2013go}.

\section{Conclusion and Outlook}
We have introduced a set of noise-robust EPR steering inequalities in high dimensions, specially designed for single-outcome detectors, that showcase the same loss-tolerance as their multi-outcome counterparts. 
We report the detection-loophole-free violation of our EPR steering inequality with a record-low one-sided heralding efficiency of $\eta = 0.038$ and a noise mixing parameter of $p^{exp} = 0.641$, which are equivalent to the loss in a 79~km-long telecommunication fibre and $35.9\%$ white noise. 
We are able to achieve this level of noise and loss tolerance by harnessing spatial entanglement in Hilbert space dimensions up to $d = 53$.
The amount of loss and noise that these inequalities can tolerate can be increased substantially by increasing the entanglement dimension $d$.
For instance, in $d = 499$, the maximum tolerable loss to demonstrate steering increases to 27~dB, which is equivalent to the loss in a 135~km-long telecom optical fibre. 
Similarly, the maximum amount of white noise that can be allowed in the system while violating the steering inequality at unit heralding efficiency reaches $95\%$. 
Finally, we demonstrate that the total measurement time required to demonstrate steering can be significantly reduced through the use of high dimensions.
For a fixed amount of noise and loss, the total integration time required to violate our steering inequality by ten standard deviations can be lowered 60 times through a modest increase to the Hilbert space dimension ($d=23$ to $d=41$).

By demonstrating resource-efficient quantum steering through realistic conditions of loss and noise, our work makes significant progress towards the implementation of one-sided device-independent quantum communication protocols. Any long-distance communication channel, such as one relying on optical fibre or free-space transmission, necessarily includes loss due to light leakage or scattering and noise due to modal dispersion or atmospheric turbulence. Our noise-robust and loss-tolerant steering inequalities demonstrate a way to overcome these detrimental effects through the use of high-dimensional entanglement. It is important to note that our methods, in particular our single-detector steering inequality, are not limited to the spatial degree-of-freedom, but can be readily extended to other photonic properties such as time-frequency \cite{Kues:2019dh} or path-encoding on a photonic integrated circuit \cite{Llewellyn:2019jv}.
Our methods could also be helpful in the demonstration of quantum memory networks on photonic platforms, which currently require high-efficiency channels for transmission, thus extending the scalability limits of long-range quantum networks~\cite{JWY+19,Pompili2021}.
Our resource-efficient steering protocol fulfils the prerequisite for 1SDI quantum key distribution (QKD)~\cite{BCW+12}, as well as private quantum computing and related protocols~\cite{Fitzsimons2017}, where establishing security between malicious servers and clients is a necessary condition. Above all else, our demonstration of quantum steering under prohibitive conditions of noise and loss shows that the fundamental phenomenon of entanglement can indeed transcend the limits imposed by a realistic environment, when one makes full use of the inherently high-dimensional photonic Hilbert space.

\begin{acknowledgements}
  This work was made possible by financial support from the QuantERA ERA-NET Co-fund (FWF Project I3773-N36), the UK Engineering and Physical Sciences Research Council (EPSRC) (EP/P024114/1), and the European Research Council (ERC) Starting grant PIQUaNT (950402).
  N.B., S.D., and R.U.~acknowledge financial support from the Swiss National Science Foundation (project 2000021\_192244/1, Ambizione PZ00P2-202179 and NCCR QSIT).
\end{acknowledgements}

\bibliography{main}
\bibliographystyle{vats}
\clearpage
\appendix
\section{Details for \texorpdfstring{$\tbeta$ and $\beta^Q(\eta, p)$}{beta values}}
\label{methods:betaLHSbin}

First, we compute an upper bound  $\tbeta$ on the LHS bound $\tbetaLHS$ for the Bob's binarised measurements defined in Eq.~\eqref{eqn:bobbin}. 
We begin with the general definition of LHS bound given in~\cite{CS16b}

\begin{equation}
\label{eqn:tbetaLHS}
  \tbetaLHS= \max_{\tmu} \Big\| \sum_{\ta,\tx}\tilde{D}_{\tmu}(\ta|\tx)\tF_{\ta|\tx}\Big\|_{\infty},
\end{equation}
which can be rewritten as

\begin{equation}
  \tbetaLHS=\max_{\tmu} \tbetaLHS_{\tmu}\quad\text{where}\quad\tbetaLHS_{\tmu}\equiv\Big\| \sum_{\tx} F_{\tmu_{\tx}|\tx} \Big\|_{\infty}.
\end{equation}
For a given deterministic strategy $\tmu$ we write $\tx\in\tmu$ for all $\tx=(a,x)$ such that $\tmu_{\tx}=1$ and the complement (corresponding to $\tmu_{\tx}=0$) will be $\tx\not\in\tmu$.
Following this notation, $|\tmu|$ will be the number of $\tx$ such that $\tx\in\tmu$.
Using the definition of Eq.~\eqref{eqn:bobbin} we get

\begin{align}
  \!\!\!\tbetaLHS_{\tmu}&=\Big\| \sum_{\tx\in\tmu} A_{a|x} +c\sum_{\tx\not\in\tmu} (\openone-A_{a|x}) \Big\|_{\infty}\\
  &= c\,\big[m(d-1)-|\tmu|\big] + (1+c)\Big\|\sum_{\tx\in\tmu} A_{a|x} \Big\|_{\infty}.\label{eqn:lastbeta}
\end{align}
The expression in Eq.~\eqref{eqn:lastbeta} is true using the fact that for any positive semidefinite $\alpha\openone+B$, where $\alpha$ is real (not necessarily positive) constant and $B$ is positive semi-definite with eigen-decomposition $\sum_j \lambda_j \ketbra{\lambda_j}$, can be written as 

\begin{align}
\|\alpha \openone + B\|_{\infty}  &=\max_j \lambda_j+\alpha,
\end{align}

since the eigenvalues $\lambda_j+\alpha$ are positive by assumption.

\noindent To continue the computation after Eq.~\eqref{eqn:lastbeta}, we make use of the bound

\begin{equation}
  \Big\|\sum_{\tx\in\tmu} A_{a|x}\Big\|_{\infty}\leq 1+\frac{|\tmu|-1}{\sqrt{d}},
\end{equation}
which is similar to results given in~\cite{Kit97,TFKW13,SC15} though it becomes increasingly loose particularly for $|\tmu|>m$.
We then get

\begin{equation}
  \tbetaLHS_{\tmu}\leq\frac{(1+c)(\sqrt{d}-1)}{\sqrt{d}}+cm(d-1)+\frac{1-c\,(\sqrt{d}-1)}{\sqrt{d}}|\tmu|,
\end{equation}
so that there is a natural choice of $c=\frac{1}{\sqrt{d}-1}.$ that allows to get an upper bound independent from  $\tmu$.
With this value, we eventually obtain the desired bound

\begin{equation}
  \tbetaLHS\leq1+m(\sqrt{d}+1)\equiv \tbeta.
  \label{eqn:desiredbLHS}
\end{equation}

Next, we evaluate the quantum value $\beta^Q$ of the functional in Eq.~\eqref{eqn:generalineq} for binarised measurements of Alice and Bob performed on a shared maximally entangled state influenced by isotropic noise $\rhoAB(p)$ in terms of one-sided heralding efficiency $\eta$, mixing parameter $p$ and we get  

\begin{align}
  \beta^Q(\eta,p)&=m\bigg\{\frac{1-p}{d}\Big[\eta+(d-\eta)\left(\sqrt{d}+1\right)\Big]\nonumber\\
  &\hphantom{=m\bigg\{}+p\Big(\eta+\sqrt{d}+1\Big)\bigg\},
  \label{eqn:betaall}
\end{align}
where $d$ is the dimension of the Hilbert space and $m$ is the number of MUB measurement settings out of $d+1$ MUBs given for prime dimension $d$~\cite{WF89}. 

With $\beta^Q(\eta, p)$ and $\tbeta$, we can simply calculate $\Delta \beta$ (see Eq.~\eqref{eqn:deltabeta}), from where we can check that to violate the steering inequality, $\eta$ must be greater than a threshold $\eta^{cr}$,

\begin{equation}
    \eta > \frac{1}{m\big(p - \frac{1-p}{\sqrt{d}}\big)} \equiv \eta^{cr}.\label{eqn:generaletacr}
\end{equation}
For a pure maximally entangled state ($p = 1$), the above threshold reduces to simply $\eta^{cr} = \tfrac{1}{m}$.

\section{Experimental details}
\label{methods:experiments}
%%%%%%%% optimising the mask explain from JTMA
\begin{figure}[h!]
  \centering
  \includegraphics[width=0.9\columnwidth]{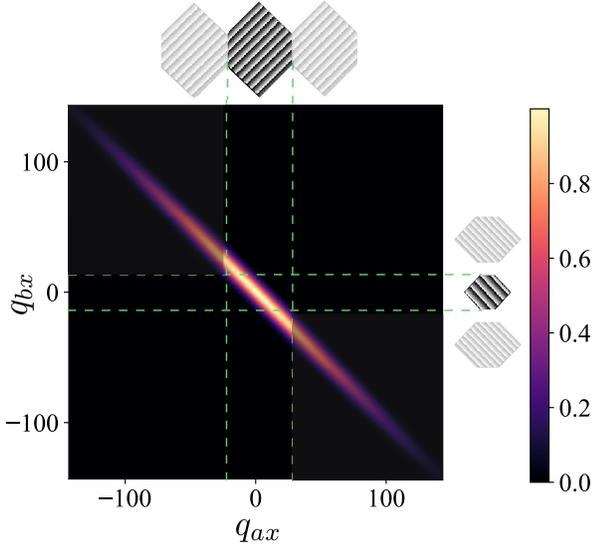}
  \caption{\textbf{Optimisation of one-sided heralding efficiency.} The one-sided heralding efficiency, or the probability that detecting a photon at the trusted party (Bob) heralds the presence of a photon at the untrusted party (Alice), can be optimised from knowledge of the two-photon joint-transverse-momentum-amplitude (JTMA)~\cite{Srivastav2021}. The size of Bob's hex-pixels can be set in such a manner that the probability of coincidence counts between Alice and Bob is increased while the probability of single counts at Bob is decreased, effectively increasing the one-sided heralding efficiency $\eta$ at Alice.}
  \label{fig:JTMAheralding}
\end{figure}
As shown in Ref~\cite{Srivastav2021}, Bob's projective measurements can be optimised from the prior information of the joint-transverse-momentum-amplitude (JTMA), such that the resultant one-sided heralding efficiency for Alice increases.
In our experiment, we employ the same optimisation to tailor the pixel masks used for Bob's projective measurements, resulting in Bob's pixel sizes being smaller than that of Alice (see Fig.~\ref{fig:JTMAheralding}). 
Additionally, the spacing between the macro-pixels in Bob's mask is chosen such that the coincidence counts between different modes (cross-talk) are suppressed (see Fig.~\ref{fig:expsetup}(a)).

%%% Detail about MUBs
The holograms on SLM on Alice and Bob perform projective MUB measurements $A_{a|x} = \ketbra{\varphi_a^x}$ on the entangled photons. They are designed according to the prescription given in~\cite{WF89}

\begin{align}\label{eqn:MUBformula}
     \ket{\varphi_a^x} = \frac{1}{\sqrt{d}}\sum_{l=0}^{d-1}\omega^{al+xl^2}\ket{l},
\end{align}
with $\omega=\exp(2\pi\mathrm{i}/d)$ is a $d$-th root of the unity.
The first basis is the computational one (individual pixel-mode), denoted $\{\ket{l}\}_{l=0}^{d-1}$, and the other $d$ bases are $\{\mub{x}{a}\}_{a=0}^{d-1}$, labelled by $x=0\ldots d-1$.

%%%% to evaluate steering inequality 
To evaluate our steering inequality in the experiment, we use coincidence counts (when both Alice's and Bob's detectors click simultaneously) and Bob's exclusive single counts (when only Bob's detectors click while Alice measures no clicks). 
For $x$th basis outcome/projector $a$ and $b$ on Alice and Bob, the coincidence counts ($C^x_{ab}$) and exclusive single counts on Bob's side ($S^x_{ab}$) are given as,

\begin{align}
  C^x_{ab}:=N^a_x \tr[A_{1|ax}\otimes \Pi_{b|x}\rho^{AB}] \label{eqn:coincidencs}\\
   S^x_{ab}:=N^a_x \tr[A_{0|ax}\otimes \Pi_{b|x}\rho^{AB}], \label{eqn:excSingles}
\end{align}
where $N^a_x $ is the total count measured by Bob. Hence, it must satisfy $ N^x_a = \sum_b ( C^x_{ab} +  S^x_{ab})$. 

Bob can then normalise the data by,

\begin{align}
  \tilde C^x_{ab}:= C^x_{ab}/ N^x_a\\
  \tilde S^x_{ab}:= S^x_{ab}/ N^x_a.
\end{align}
to obtain the steering inequality elements,

\begin{align}
  \tr [ \tF_{\ta|\tx} \sigma_{\ta|\tx}] =\begin{cases}
    \tilde C^x_{aa},& \ta=1\\
    c \sum_{b\neq a}\tilde S^x_{ab}, &\ta=0
  \end{cases}\label{eqn:functinalCCexcS}
\end{align}
and the inequality can be evaluated by

\begin{align}
  \beta^Q = \sum_{a, x} \left( \tilde C^x_{aa}+c \sum_{b\neq a} \tilde S^x_{ab} \right).
\end{align}
Similarly, we calculate the one-sided heralding efficiency $\eta^{exp}$ for $x$-th measurement setting on Alice's channel from

\begin{equation}
    \eta^{exp} = \sum_{a,b}\tilde C^x_{ab}.
\end{equation}\label{eqn:onesidedefficiency}
Note that in our work we perform only $m = d$ MUB measurements excluding the computational basis. 
The one-sided heralding efficiency $\eta^{exp}$ for each MUB measurement (not the computational basis) in dimensions $d$ does not vary significantly. To characterise the level of noise in the system, we use the amount of cross-talk $v$ between pixel modes:

\begin{equation}\label{eqn:v}
    v = \frac{\sum_a \tilde C^x_{aa}}{\sum_{ab} \tilde C^x_{ab}}.
\end{equation}
This estimate is valid because the $v$ does not change substantially across the MUBs and thus, it behaves isotropically. Formally, the mixing parameter $p^{exp}$ in the experiment is given as

\begin{equation}\label{eqn:mixingparameterexp}
    p^{exp} = \frac{vd-1}{d-1}.
\end{equation}

\section{Minimising total measurement time \texorpdfstring{$T$}{T}}
\label{methods:confidence}

The expectation (mean) number of coincidences and exclusive single counts at Bob's side are,

\begin{align}
 \av{ C^x_{ab}} &= N \eta \left( \frac{p \delta_{ab}}{d}+ \frac{1-p}{d^2}\right),\\
\av{ S^x_{ab}} &= \frac{N }{d}-C^x_{ab},
\end{align}
where $N=R t^{ac}$ is the total number of copies of the state Bob receives during an acquisition window, $t^{ac}$, given the underlying single count rate $R$ detected at Bob's side. In our experiment, we assume that the single count rates $R$ does not vary significantly for different dimensions. 
Since the statistics of the raw counts are Poissonian, their variances are, $\text{Var}( C^x_{ab}) = \av{ C^x_{ab}}$ and $\text{Var}( S^x_{ab}) = \av{ S^x_{ab}}$.
We can estimate the variance of $\beta^Q$ by,

\begin{align}
  \text{Var}(\beta^Q) = \sum_{x a b} \left[ \left(\frac{\partial  \beta}{\partial  C^x_{ab}}\right)^2\text{Var}( C^x_{ab})+\left(\frac{\partial  \beta}{\partial  S^x_{ab}}\right)^2\text{Var}(S^x_{ab}) \right]
\end{align}
which is inversely proportional to $N$ (and therefore $R$ and $t^{ac}$), so we can factor out,

\begin{equation}\label{eqn:C4}
    \text{Var}(\beta^Q)=\frac{f(\eta,p,d,m)}{N},
\end{equation}
where $f(\eta,p,d,m)$ is a function of $\eta,p,d$ and $m$  which is independent of $N$ (equivalent to $R$ and $t^{ac}$).
We wish to find the dimension that minimises the total experiment time, $T= t^{ac} m d^2$, whilst violating the steering inequality by 10 standard deviations. For an expected violation $\Delta \beta$, we require,

\begin{align}\label{eqn:C5}
  \Delta \beta \geq 10 \sqrt{ \text{Var}( \beta^Q)}
\end{align}
From Eq.~\eqref{eqn:C4} and \eqref{eqn:C5}, we can then solve for $N$,

\begin{align}
  N\geq  10^2\frac{f(\eta,p,d,m)}{(\Delta \beta)^2},
  \label{eqn:Nsol}
\end{align}
which is valid only when $\Delta \beta >0$.
We then evaluate the total measurement time to saturate the bound,

\begin{align}
  T = \frac{Nmd^2}{R}= \frac{m d^2 10^2}{R} \frac{f(\eta,p,d,m)}{(\Delta \beta)^2}.
  \label{eqn:timevsDim}
\end{align}
Interestingly, at fixed value of $\eta$ and $p$, for $m = d$ measurement settings, the expression $\tfrac{m d^2 f(\eta,p,d,m)}{(\Delta \beta)^2}$ depends  non-monotonically on dimension $d$. 
This makes the total measurement time $T$ to reach a minimum at a non-trivial dimension $d$.
The plot in Fig.~\ref{fig:timevsDim} shows the dependence of $T$ with respect to $d$ at fixed heralding efficiency $\eta= 0.062$ and noise level $p=0.775$ and rate $R\sim 10^5$. 
\end{document}